# Ground State Rotational Bands of Deformed *e-e* Nuclei

Mohamed E. Kelabi[*]
A. Y. Ahmed[*]
Vikram Singh[†]

**Abstract**
In the frame work of the hydrodynamical model, a new model of the ground state rotational bands of deformed *e-e* nuclei is developed by introducing the variable moment of inertia, and the effect of $\beta$- and $\gamma$- vibrational bands. The model is applied to calculate the energies of the ground state band of $^{158}Dy$. The results of our calculations are in close agreement with data compared with other existing models.

**Introduction**
Recently Abzouzy and Antony[1] have shown that the semi-empirical expression of Sood[2] works well for *Th* and *U* nuclei, which is obtained from the power series of angular momentum factor $I(I+1)$ as

$$\frac{C}{B} = \frac{D}{C} = \cdots = N\frac{B}{A} \qquad (1)$$

where *A*, *B*, *C*, *D* …. are the coefficients of successive order correction terms[3]. They have then assigned *N* to depend on two parameters

$$N = a - bI \qquad (2)$$

and deduced their values[1]: *a* = 2.55 and *b* = 0.05 for the best fit.

**Theory and formalism**
A more systematic calculations can be carried out by considering only the Rotational-Vibrational model RVM of Bohr and Mottelson[4]. This was done by adding the effect of nuclear rotation[2] on the moment of inertia, thus one writes

$$E(J) = A(J)J(J+1) - B(J)J^2(J+1)^2 \qquad (3)$$

where

$$A(J) = \frac{\hbar^2}{2\mathcal{J}(J)} \qquad (4)$$

---

[*] Physics Department, Al-Fateh University, Tripoli, LIBYA
[†] B-1/1051, Vasant Kunj, New Delhi 110070.

$$B(J) = \frac{4A(J)^3}{(\hbar\omega_\gamma)^2} + \frac{12A(J)^3}{(\hbar\omega_\beta)^2} \qquad (5)$$

here $\hbar\omega_\beta$ and $\hbar\omega_\gamma$ are the head energies of the $\beta$- and $\gamma$-vibrations[5], respectively. For convenience, we rewrite Eq. (3) as

$$E(J) = A(J)J(J+1)\left[1 - \frac{B(J)}{A(J)}J(J+1)\right] \qquad (6)$$

where the ratio $\frac{B(J)}{A(J)}$ takes the form

$$\frac{B(J)}{A(J)} = \frac{\hbar^4}{\mathcal{J}(J)^2}\left[\frac{1}{(\hbar\omega_\gamma)^2} + \frac{3}{(\hbar\omega_\beta)^2}\right]. \qquad (7)$$

In order to write $\mathcal{J}(J)^{-1}$ and $\mathcal{J}(J)^{-2}$ in terms of the ground state value $\mathcal{J}_0$[6], which corresponds to $J = 0$, we make Taylor series expansion, giving

$$\frac{1}{\mathcal{J}(J)} = \frac{1}{\mathcal{J}_0}\frac{1}{1+\sigma_1 J + \sigma_2 J^2 + \cdots} \qquad (8)$$

$$\frac{1}{\mathcal{J}(J)^2} = \frac{1}{\mathcal{J}_0^2}\left(1 - 2\sigma_1 J - 2\sigma_2 J^2 + 3\sigma_1^2 J^2 + 6\sigma_1\sigma_2 J^3 - \cdots\right) \qquad (9)$$

where, we have introduced the Softness parameter[7], defined by

$$\sigma_n = \frac{1}{n!}\frac{1}{\mathcal{J}_0}\frac{\partial^n \mathcal{J}(J)}{\partial J^n}\bigg|_{J=0}. \qquad (10)$$

Keeping only the first order terms in $\sigma_n$, Eqs. (8) and (9) become

$$\frac{1}{\mathcal{J}(J)} = \frac{1}{\mathcal{J}_0}\frac{1}{1+\sigma_1 J} \qquad (11)$$

$$\frac{1}{\mathcal{J}(J)^2} = \frac{1}{\mathcal{J}_0^2}(1 - 2\sigma_1 J). \qquad (12)$$

Inserting Eqs. (11) and (12) into (4) and (7), respectively, and rearranging Eq. (6), one easily obtains

$$E(J) = \frac{A_0}{1+\sigma_1 J}J(J+1)\left[1 - \frac{B_0}{A_0}(1-2\sigma_1 J)J(J+1)\right] \qquad (13)$$

which can be simplified to give

$$E(J) = \frac{A_0}{1+\sigma_1 J} J(J+1) - cA_0^3 \frac{(1-2\sigma_1 J)}{1+\sigma_1 J} J^2(J+1)^2 \qquad (14)$$

where, $A_0 = \frac{\hbar^2}{2\mathcal{J}_0}$ and $B_0 = cA_0^3 = \left[\frac{4}{(\hbar\omega_\gamma)^2} + \frac{12}{(\hbar\omega_\beta)^2}\right] A_0^3$.

This third order nonlinear expression, Eq. (14), can be used to calculate the ground state rotational bands of deformed *e-e* nuclei.

**Results**

We have applied Eq. (14) to calculate the energies of the ground state band of $^{158}Dy$. In these calculations, we have treated $A_0$, $\sigma_1$ and $c$ as three free parameters, which can be determined by fitting the first excited states $2^+$, $4^+$, and $6^+$ with the available data. In Fig. 1 we schematically present the results of our calculations along with the experimentally observed energies and the prediction of other existing models[8]. The results of our model are in close agreement with data than the other models.

$A_0 = 17.042$ KeV    $\sigma_1 = 13.806$    $c = 3.440 \times 10^{-6}$ KeV$^{-2}$

| $J^\pi$ | Exp | VMI | VAVM | GVMI | Present work |
|---|---|---|---|---|---|
| $18^+$ | 3781.7 | 3943.0 | 3913.3 | 4030.3 | 3866.3 |
| $16^+$ | 3190.7 | 3263.8 | 3248.5 | 3329.3 | 3221.0 |
| $14^+$ | 2612.6 | 2628.6 | 2623.7 | 2674.9 | 2613.3 |
| $12^+$ | 2049.2 | 2041.7 | 2043.5 | 2071.9 | 2043.1 |
| $10^+$ | 1519.9 | 1508.8 | 1513.5 | 1526.2 | 1516.2 |
| $8^+$ | 1044.1 | 1036.9 | 1041.4 | 1045.0 | 1043.1 |
| $6^+$ | 637.87 | 637.87 | 637.87 | 637.87 | 637.87 |
| $4^+$ | 317.26 | 317.26 | 317.26 | 317.26 | 317.26 |
| $2^+$ | 98.94 | 98.94 | 98.94 | 98.94 | 98.94 |
| $0^+$ | 0.0 | 0.0 | 0.0 | 0.0 | 0.0 |

Fig.1. Experimental and calculated energy levels of $^{158}Dy$ in [KeV].

**Conclusion**

The present model Eq. (14) is practically fit to predict the ground state rotational bands of almost all deformed *e-e* nuclei, and can also be applied to nuclei where the energies of *β*- and *γ*- vibrations are experimentally available. With the known values of $\hbar\omega_\beta$ and $\hbar\omega_\gamma$, contained in parameter *c* of Eq. (14), the model reduces to only two parametric expression, hence, the higher spin states can be evaluated by fitting only the first excited states $2^+$ and $4^+$ with experimental data.